\documentclass[sigconf]{acmart}

\copyrightyear{2021}
\acmYear{2021}
\setcopyright{acmcopyright}\acmConference[WiPSCE '21]{The 16th Workshop in Primary and Secondary Computing Education}{October 18--20, 2021}{Virtual Event, Germany}
\acmBooktitle{The 16th Workshop in Primary and Secondary Computing Education (WiPSCE '21), October 18--20, 2021, Virtual Event, Germany}
\acmPrice{15.00}
\acmDOI{10.1145/3481312.3481347}
\acmISBN{978-1-4503-8571-8/21/10}

\usepackage[tight,footnotesize]{subfigure}
\usepackage{booktabs} 
\usepackage{todonotes}
\usepackage{microtype}
\usepackage{xspace}
\usepackage{balance}
\usepackage[utf8]{inputenc}
\usepackage[ngerman,english]{babel}

\usepackage[compact]{titlesec}
\usepackage{fontawesome}

\begin{document}

\title{An Experience of Introducing Primary School Children to Programming using Ozobots (Practical Report)}


\settopmatter{authorsperrow=3}

\newcommand{\passau}{Passau\xspace}
\newcommand{\interreg}{INTERREG\xspace}
\newcommand{\countries}{Germany and Austria\xspace}
\newcommand{\german}{German\xspace}


\author{Nina Körber}
\affiliation{%
 \institution{University of Passau}
 \city{Passau}
 \country{Germany}
}
\author{Lisa Bailey}
\affiliation{%
 \institution{University of Passau}
 \city{Passau}
 \country{Germany}
}
\author{Luisa Greifenstein}
\affiliation{%
 \institution{University of Passau}
 \city{Passau}
 \country{Germany}
}
\author{Gordon Fraser}
\affiliation{%
 \institution{University of Passau}
 \city{Passau}
 \country{Germany}
}
\author{Marina Rottenhofer}
\affiliation{%
 \institution{Johannes Kepler University Linz}
 \city{Linz}
 \country{Austria}
}
\author{Barbara Sabitzer}
\affiliation{%
 \institution{Johannes Kepler University Linz}
 \city{Linz}
 \country{Austria}
}

\begin{abstract}
	Algorithmic thinking is a central concept in the context of computational
thinking, and it is commonly taught by computer programming. A recent trend
is to introduce basic programming concepts already very early on at primary
school level. There are, however, several challenges in teaching programming at
this level: Schools and teachers are often neither equipped nor trained
appropriately, and the best way to move from initial ``unplugged'' activities
to creating programs on a computer is still a matter of open debate. 
In this paper, we describe our experience of a 
project aiming at  
supporting local primary schools in introducing children to programming concepts using Ozobot robots.
These robots have two distinct advantages: First, they can be programmed with and without
computers, thus helping the transition from unplugged programming to programming
with a computer. Second, they are small and easy to transport, even when used
together with tablet computers. Although we learned in our outreach events that
the use of Ozobots is not without challenges, our overall experience is
positive and can hopefully support others in setting up first encounters with
programming at primary schools.
\end{abstract}

\ccsdesc[500]{Social and professional topics~Software engineering education}
\ccsdesc[500]{Social and professional topics~K-12 education}

\keywords{Ozobots, Programming Education, Primary School Programming} 

\maketitle

\section{Introduction}

Computational thinking~\cite{wing2006computational} (CT) is an important aspect of education, and programming is a primary method in order to teach it~\cite{lye2014review,geldreich2018off}. Establishing an elementary understanding of algorithmic thinking early on is important to successfully achieve computational thinking, but also to counter common preconceptions against programming and computer science that influence the gender imbalance later on~\cite{metz2007attracting,madill2007developing,steele1997threat,sullivan2013gender}. Programming is also increasingly used as an instrument to teach other topics. The best methods for teaching programming early on, however, are still being explored.

A common approach to introducing children to programming concepts is by using ``unplugged'' activities that require no computers~\cite{bell2009computer,faber2017teaching}, for example by letting the children impersonate robots executing sequences of instructions. A challenging next step is to transfer concepts learned with unplugged programming to programming with computers. Physical computing with programmable devices such as robots is a possible intermediary step, as they have a tangible, real-world impact through their actions~\cite{zaharija2013introducing,demo2008concrete}.
However, this requires schools to have access to appropriate hardware, and teachers need to have the confidence and knowledge to manage hardware devices and to support children in using them.

While CT is already integrated into curricula in many countries,
primary schools in \countries currently do not touch upon the topic. In an endeavour to help local primary schools with the introduction of programming concepts, we therefore visited eight primary schools from October 2018 to October 2019. During these school visits, we used the Ozobot Evo\footnote{https://ozobot.com/, last accessed June 1, 2021} programmable
robots in two-hour workshops. Ozobots are small programmable robots for which
initial experience reports are generally
positive~\cite{fojtik2017ozobot,van2018best,vzavcek2019development}.
The Ozobots have several advantages: They bridge the gap between unplugged
programming and physical programming, as they can be used with pen and
paper as well as on a tablet or desktop computer; they are very small and
portable, and can easily be moved from one school to another;
the programming environment is very flexible and the level of difficulty can be adapted from pre-reading stage up to complex
programming concepts.

In this paper, we describe the workshop format we developed for our outreach activity, and share our generally positive experience of interacting with children and Ozobots. Overall, the children were enthusiastic as one would expect, both when interacting with pen and paper as well as with tablet computers, although the Ozobots proved to be better suited for their pen-and-paper mode.

\section{The Ozobot Evo Robot}

\begin{figure}[tb]
	\centering
    \vspace{2em}
	\includegraphics[width=0.6\columnwidth]{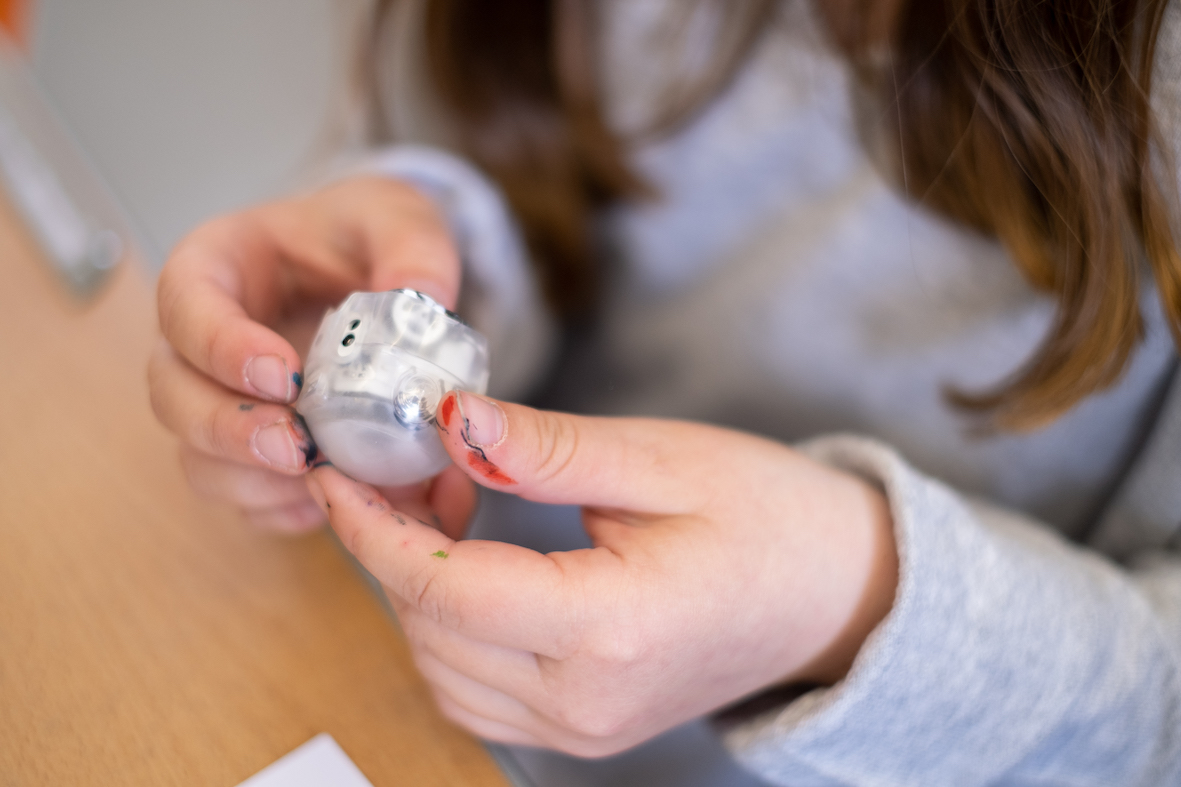}
    \vspace{-1em}
	\caption{\label{fig:bot}The Ozobot Evo robot: Small, easy to handle, with a soft plastic cover that makes it sturdy. The colourful fingers of the child in the picture are a common result of the pen-and-paper mode of interaction.}
    \vspace{-1em}
\end{figure}

Figure~\ref{fig:bot} shows an Ozobot Evo robot. The robot can make sounds, has multiple lights that can flash in different colours, and it can move. It has sensors to detect the colour of the surface it is driving on, and proximity sensors facing to the front and back.

\begin{figure}[t]
	\centering
	\includegraphics[width=0.6\columnwidth]{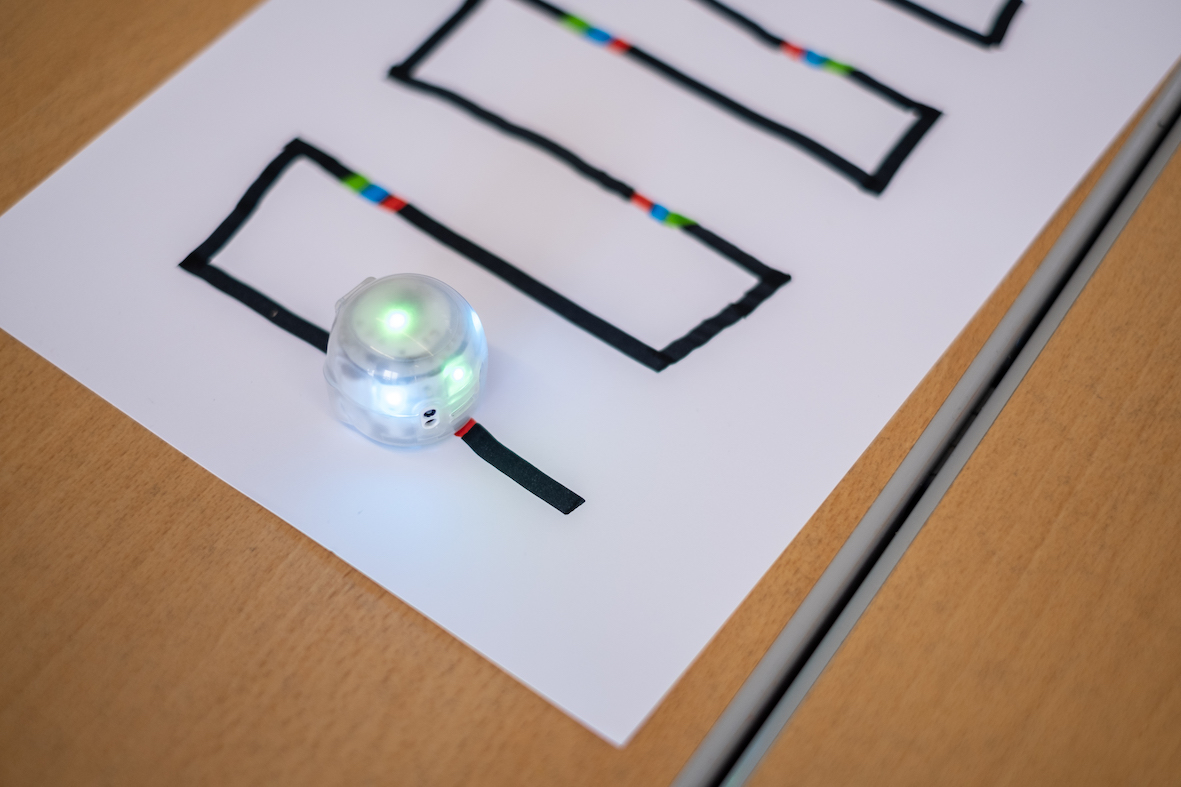}
    \vspace{-1em}
	\caption{\label{fig:malen}Ozobot robot moving along a line with embedded colour codes. In this example, green-blue-red increases the counter by one (the symmetric colour sequence red-blue-green decreases the counter).}
    \vspace{-1em}
\end{figure}

\begin{figure}[t]
	\centering
	\includegraphics[width=0.6\columnwidth]{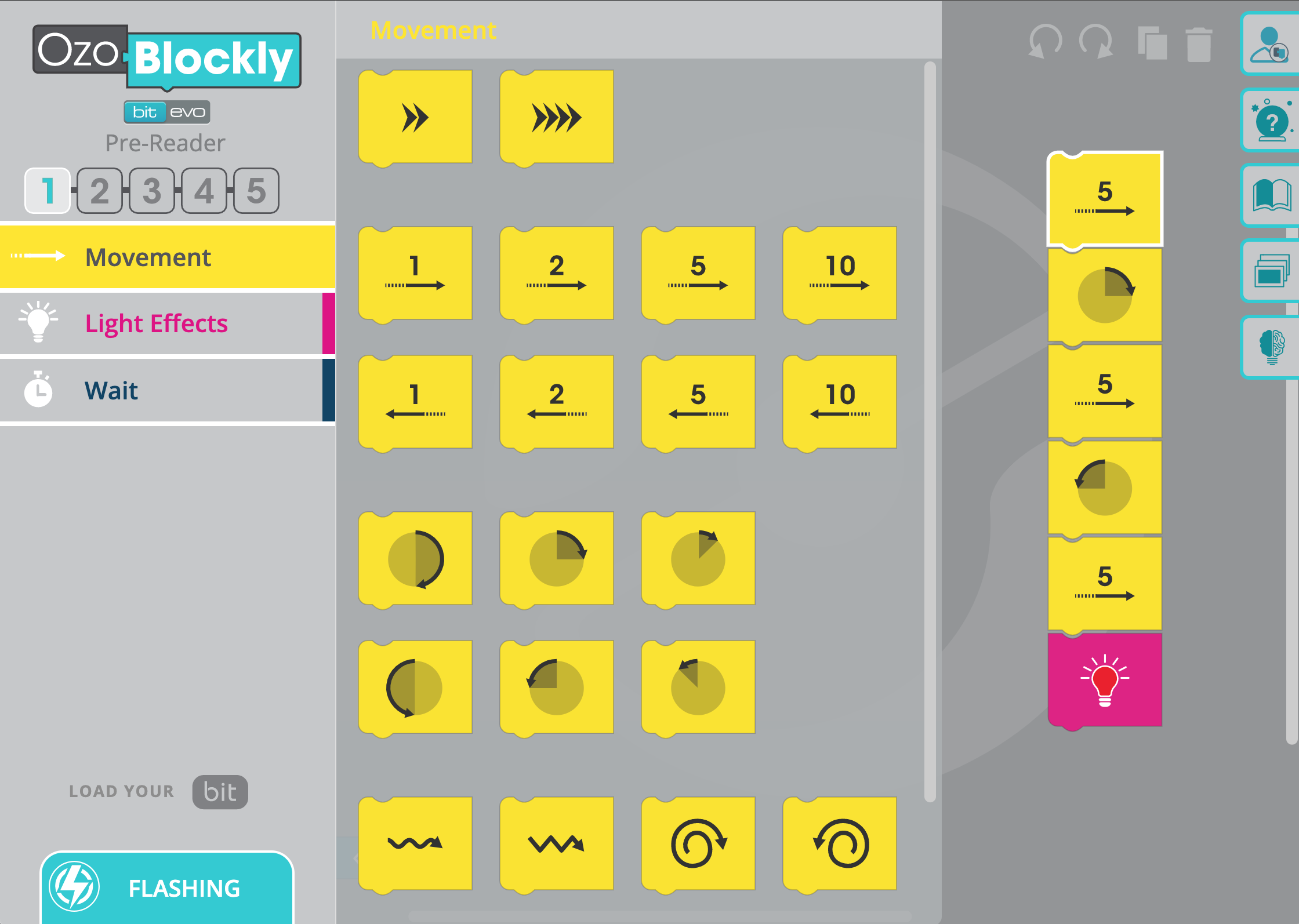}
    \vspace{-1em}
	\caption{\label{fig:ozoblockly}The OzoBlockly programming environment in a web browser: Categories of blocks are shown on the left, and clicking on a category opens a drawer of blocks available. At level 1 (``Pre-Reader'') only a few blocks are shown, but many more blocks are available at increasing levels.}
    \vspace{-1em}
\end{figure}

There are two main ways of interacting with an Ozobot: In the line-following mode, the robot follows coloured lines it senses with its bottom sensors. While doing so, it changes its lights based on the colour of the line it is following. Lines may contain special colour sequences, which represent commands (e.g., turn left, turn right, go fast, dance, etc.). Figure~\ref{fig:malen} shows an example of the robot increasing a counter while following a line. The main categories of commands are related to speed (from snail dose to nitro boost), special moves (spinning, zigzagging, etc.), directional movement (turning, jumping lines, u-turns, etc.), counters (e.g., counting colours, counting crossings, increasing/decreasing counters), timers, and commands to indicate success or completion of a task\footnote{https://play.ozobot.com/print/guides/ozobot-ozocodes-reference.pdf, last accessed June 1, 2021}.

Ozobots can also be programmed via the OzoBlockly app\footnote{https://ozobot.com/create/ozoblockly, last accessed June 1, 2021} (Figure~\ref{fig:ozoblockly}). The programming environment follows the standard block-based approach with a toolbox of different categories of blocks on the left hand side of the screen.  OzoBlockly supports five different levels, where in the easiest level blocks are purely graphical with icons and only a few numbers instead of text, so that programming can be done even at pre-reading stages. The basic commands are reminiscent of the Logo~\cite{logo} programming approach, and mainly control the movement of the robots. At the highest level, the programming language supports even complex constructs such as functions with return values. OzoBlockly programs can be transferred to the robots by ``flashing'', which means holding the robot face down on the screen while the program is transferred by a colour sequence. When using a tablet rather than a desktop computer, the Ozobot Evo can also be controlled via an app and a bluetooth connection. The app contains not only OzoBlockly, but also various functionalities such as a remote control. 



\section{Design of the Workshop Format}
In this section, we describe the theoretical underpinnings that informed the design of our outreach programming workshops. 
When it comes to the selection of appropriate teaching material, the target group and the learning objectives but also the desired learning approach must be kept in mind. Our outreach activities aimed at the children programming a sequence and achieving a positive attitude towards programming. We used a step-by-step approach combined with active involvement of the children. This is suitable for primary school children as they are not overwhelmed but experience success when discovering new aspects of programming the Ozobot. We decided to base our activities on a selection of worksheets found online\footnote{https://ozobot.com/educate/lessons, last accessed June 1, 2021} as they fit our desired approach as follows.

\subsection{Cognitive Load Theory}
A renewed conception of the Cognitive Load Theory distinguishes between intrinsic and extraneous cognitive load \cite{sweller2010element}. Both are caused by the characteristics of the learning task and the learner and deal with the resources of the working memory and the long-term memory \cite{choi2014effects}. The extraneous cognitive load is high when the instruction is not optimized and thus distracting; as this blocks cognitive capacities without positive effects in learning, it must be minimized \cite{sweller2010element}. Ozobot worksheets for the line-following mode 
cause low extraneous cognitive load: They are clearly adapted to the learning objective, similar to each other, simple in their structure and therefore not distracting. This is because all worksheets contain one to two paths with gaps and the children have to fill them with either a line or a colour code. The remaining cognitive resources are available for dealing with the intrinsic cognitive load. As this should neither be too high nor too low, each worksheet has one to two clearly defined and achievable learning objectives.

\subsection{Use-Modify-Create Framework}


We designed activities to proceed according to the use-modify-create framework~\cite{lee2011computational}: At first the children \emph{use} given lines or colour codes
, then \emph{modify} the path by inserting self selected colour codes 
and at last \emph{create} an own path
. When introducing programming with the OzoBlockly app the children first \emph{use} some given sequences of commands by executing them in an unplugged approach. The children can also be tasked to recreate and in this way \emph{use} an incorrect implementation within the app. The incorrect script must then be \emph{modified} and at last the children can \emph{create} their own program
.

\section{Programming Workshop Format}

Our outreach activity was organised in the form of two-hour workshops at primary schools, targeting children in grades 3--5. We visited eight schools between October 2018 and October 2019. Each workshop was supervised by 2--4 project participants, and the class sizes were around 20--25 students (sometimes covering two classes at once). 
In this section, we describe the sequence of activities conducted at the workshops. 
These activities are based on teaching material available online\footnote{https://ozobot.com/educate/lessons, last accessed June 1, 2021} as well as our own ideas.

\subsection{Activity 1: Getting to Know the Ozobots}

After a brief introduction, we discussed with the children what a robot is, what robots they know, and what they look like. Typically, they expected robots to be large and humanoid. As we wanted the children to co-construct their insights, we gave one robot to two children
As initial activity, we asked them to find out what equipment and capabilities the robot has, and how to turn it on. After a while, we let the children tell us what they had discovered and wrote down the correct suggestions on the blackboard. This activity proved to be helpful in order to set realistic expectations; e.g., the robots do not have elastic springs and therefore are not able to jump. This was particularly useful to understand for when the children were later introduced to ``line jumping'' commands, which refer only to switching from lines to follow rather than physical jumping. Ozobots have a single button that is not entirely obvious and took the kids a while to locate. They discovered the wheels and the USB charging port quickly, and once they had managed to turn on the robot, they discovered the lights and sounds the robot can produce. The proximity and colour sensors are not obvious and needed  explaining; this immediately lead to the second activity.

\subsection{Activity 2: Line-Following}

To explain how the robot uses the colour sensor, we handed out the first worksheet\footnote{https://storage.googleapis.com/ozobot-lesson-library/3-5-basic-training-color-codes/3-5-Basic-Training-Color-Codes-full-version.pdf, p. 24, last accessed June 1, 2021}, which consists of a black line with gaps in it. The children received coloured pens and were asked to fill the gaps and observe how the robots behave. There were two learning objectives: First, the children were supposed to discover that the Ozobot follows lines, and while doing so, the lights glow in the colour that the colour sensor detects (white instead of black, as only exception). Second, the children got a feeling of how thick the lines have to be so that the robot can detect them. The first worksheet was followed by a second one\footnote{Ibid., p. 25} that also consists of lines with gaps, but this time the gaps are intentionally located at corners and bends. The very similar structure decreases the extraneous cognitive load~\cite{sweller2010element} and the children can focus on drawing appropriate lines the robot can follow. 
Children that completed the task quicker than others were allowed to use the time for extending the lines or drawing their own lines on the back side of the paper.

\subsection{Activity 3: Introduction to Commands}

A short discussion about how the colour sensor works
lead to the third activity, which is concerned with how the colour sensors can be used to control the robots. The third worksheet\footnote{Ibid., p. 26} contains a circular line with two gaps, and two colour codes that the children need to draw into the gaps. Once the children had had some time to try out these codes, we discussed with them how the two codes caused the robots to behave. Each code consists of a sequence of three colours; in this first example, both codes are symmetric, so that the robot behaves identical regardless of the direction in which it traverses the code. With the fourth worksheet\footnote{Ibid., p. 27}, the children again received a line with two gaps and two corresponding codes, but this time only one code is symmetric, while the second one has a different effect depending on the direction of traversal (in one direction it slows the robot down, in the other it accelerates it). Again, we  discussed this with the children after they had experimented with the worksheet to help them construct their own insights 
.

\subsection{Activity 4: Discovering Randomness}

The fifth worksheet\footnote{Ibid., p. 28} consists of two basically equal paths, each containing a crossing which leads to three different colours; in the second path, there is a gap for inserting a command. As first activity, we tasked the children to conduct an experiment where they start their robot ten times on the first path, and keep track of how often the robot moves to each of the colours. Once completed, we collected the data on the blackboard, to see which colour was targeted most of the times (to simplify the counting, we only asked each child or pair of children which colour was visited most frequently). Since the Ozobot makes a random choice at a crossing unless instructed otherwise, the result is different every time. After explaining that the robot chooses randomly, we handed out tables with the main colour codes that the Ozobot can handle%
, and turned to the second path on the worksheet. After explaining the different commands on the command table, we chose an arbitrary colour (usually the one that was least frequently visited during the previous experiment) and then tasked the children to select and draw a command that will make the robot go to that colour every time.

\subsection{Activity 5: Controlling the Robot}

At this point, the children had an understanding of the capabilities and the behaviour of the robots and they knew how to use colour codes as well as which commands the robot can understand. We thus handed out a challenge (sixth worksheet) from the Ozobot website. In these challenges, there typically is a scenario (e.g., the road from home to school\footnote{Ibid., p. 29}) with a clear objective (e.g., get the robot from home to school) and obstacles that require the use of colour codes (e.g., dead ends, road works, etc.) Solving such a challenge sometimes took a while, and it was useful to be prepared to handle mistakes (either by handing out a new sheet, or by using white stickers to cover incorrect commands). Children who completed the task early were allowed to create their own racetrack on the back of the worksheet which goes hand in hand with the learning progression described in the use-modify-create framework~\cite{lee2011computational}. 

\subsection{Activity 6: Introduction to Programming}

The next activity was the transition from pen-and-paper commands to sending commands via programming. This activity was done without robots (we tried to use this time for recharging the robots a bit). We introduced the commands in the OzoBlockly format using large paper printouts\footnote{https://storage.googleapis.com/ozobot-lesson-library/ozoblockly-training-k-1/ozoblockly-training-k-1.pdf, pp. 14-20, last accessed June 1, 2021} of the easiest OzoBlockly level, which consists of graphical blocks. Using these printouts, we focused on the computational thinking concept of \emph{sequences}~\cite{brennan2012new}: We pinned sequences of commands on the black board, and then asked the children to execute the programs (which typically consisted of walking, turning, and making happy/sad faces). While one child executed the program, the other children were tasked to judge if the execution was correct or not. Finally, we demonstrated the OzoBlockly app on a tablet computer, showing how the same commands can be arranged in sequences there.

\subsection{Activity 7: Programming the Ozobots}

In the final activity, each pair of children was given a tablet computer (and a recharged Ozobot, if necessary). We explained how to connect the tablet computer to the robot via bluetooth, and how to use the OzoBlockly app. Then, we gave the children the seventh worksheet consisting of the first coding challenge\footnote{Ibid., p. 13}. The robot has to go from the start position to its room in its house, and we pre-loaded the OzoBlockly app with an incorrect implementation of the sequence of instructions, leading to the robot erroneously ending up in the forest. The task for the children was to correct the program so that the robot successfully walks to its home. Consequently, the main computational thinking concept reinforced by this activity is again \emph{sequences}, as well as the computational thinking practice~\cite{brennan2012new} of \emph{testing and debugging}. Once the task was completed, the children were given the chance to be creative with OzoBlockly. Over the course of the workshops, we refined this activity to the task of programming the Ozobot to perform a dance.

\begin{figure}[t]
	\centering
	\subfigure[Starting point]{
	\includegraphics[scale=0.2]{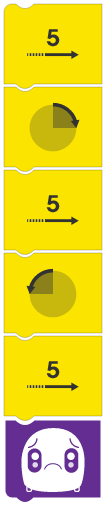}
	}
	\hfill
	\subfigure[Expected solution]{
	\includegraphics[scale=0.2]{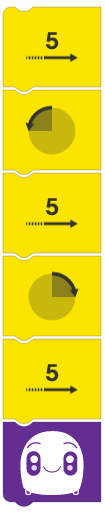}
	}
	\hfill
	\subfigure[A workshop participant conducting the programming activity]{\includegraphics[width=0.56\columnwidth]{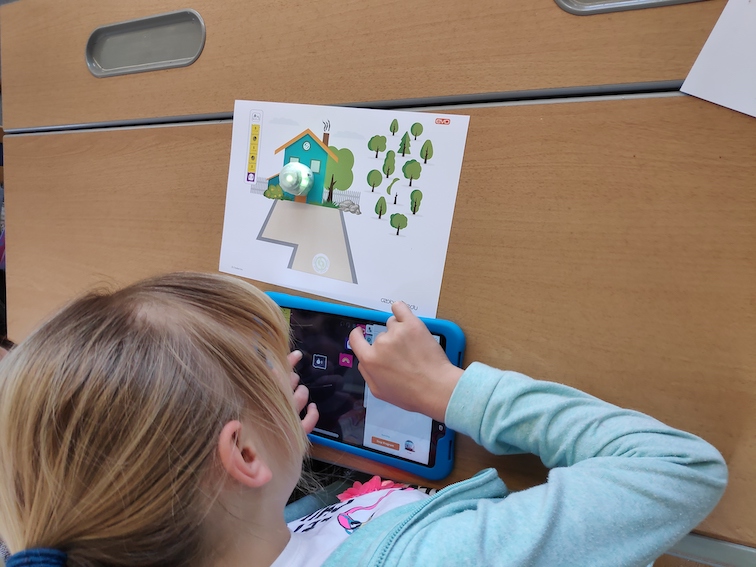}}
    \vspace{-1em}
	\caption{The first OzoBlockly programming task is to let the robot move home safely. The starting point is an incorrect program that moves the robots to the forest. The correct solution requires reordering the rotation instructions as well as replacing the emotion command at the end.} 
    \vspace{-1em}
\end{figure}



\section{Lessons Learned}

\subsection{Resources: Robots in General}
\paragraph{\bf Labeling Robots.} It is helpful to name  the robots and to attach stickers with their names. Not only for matching tablet computers with robots and connecting via bluetooth the name labels are true life savers---which was perceived similarly in a university course~\cite{mayerova2019creating}---the children also identify much stronger with their robots if they have names. This is why robots that work without tablets (e.g., BeeBots) should also be labeled. We used two different sets of robots, one labeled with numbers, and the other with real names such as ``Fred'', ``Lena'' or ``Bobo'', and in every class the favourites were the ones with names.

\paragraph{\bf Remote Control.} In some of the classes we demonstrated that the robots can be controlled using the remote control functionality in the OzoBlockly app. As it is easier and apparently more fun to use the joystick than to create programs, some children started playing with the robots rather than programming them. We recommend to hide the remote control functionality (also available for other robots, e.g. for Sphero robots in the App Sphero Play) and hope that the children do not discover it themselves. If they do so, one has to try to convince the children not to use it. For example, you can tell them that they miss out on practicing to be a cool programmer when only playing with a toy that even toddlers can handle.

\paragraph{\bf Bluetooth Kills Battery Life.} While Ozobots can easily follow lines for 90 minutes, the bluetooth connection seems to drain batteries extremely quickly~\cite{mayerova2019creating}. We found it necessary to recharge robots before switching from drawing to programming mode, and even then do the robots not last very long. Thus, we recommend to always turn off the robots during phases of discussion, and to plan for breaks where the robots can be recharged or for spare robots.

\paragraph{\bf Pairing Robots and Tablets.} It is officially recommended that each tablet computer is matched with exactly one robot. This may work for other robots, but we found this not to be a viable solution since Ozobots run out of battery too quickly when connected via bluetooth. However, when connecting different robots and tablet computers, it frequently happens that the app refuses to control the robot until it has been ``updated'' -- even if the robot is already completely up to date. Performing such an update requires a working internet connection, which may not be available during outreach activities. Also, both duration and success of these updates were rather non-deterministic, which led to some robots not being usable with tablets at all. Even though this seems to be a common problem~\cite{mayerova2019creating}, the Ozobot support was not able to recommend an alternative at the time we contacted them. 

\paragraph{\bf Correct Programs Appearing Erroneous.}
In Activity~7, sometimes the execution of a program did not succeed even though the program was correct. The reason was that in consecutive executions the robot usually did not have the exact same direction at the starting point, therefore ending at a slightly different position at every execution. However, as soon as the children had understood, they could differentiate between the usual small deviation present even for correct programs and programs which were indeed erroneous, leading to the robot obviously missing the end point.

\subsection{Resources: Ozobots}
\paragraph{\bf Pen-and-Paper Drawing.} The children are fascinated with drawing creative patterns for the robots to follow with the Ozobot markers (which can be replaced with regular markers once they are out of colour). When producing worksheets or handing out blank sheets of paper to draw on, make sure not to use standard printer paper but thicker paper, otherwise the desks will be coloured.

\paragraph{\bf Pen-and-Paper Tasks.} When given tasks where children have to fill in colour codes, mistakes will be made; these can be repaired with Ozobots~\cite{fojtik2017ozobot}. It is important to either have white stickers to allow corrections, or to have spare copies of the worksheets. In the worst case, a reasonable workaround is to draw detour lines to allow the Ozobots to bypass erroneous solutions.

\paragraph{\bf Online Resources.}
We encountered challenges with individual activities such as ``Modeling animal habits''\footnote{https://classroom.ozobot.com/lessons/lnbqFXcCmYQnSitM2dmN6a8AA1, last accessed June 1, 2021 (available after registration)}, which uses point-counter colour codes to let the Ozobot count from five to zero points, taking into account the non-orthogonal design of the crossings and which route the Ozobot will most likely take. The children understood the task well and filled the map with suitable colour codes. After counting from five to zero, the robot is supposed to stop and blink red, but it did not always do so, even for correct solutions. As this lack of instant positive feedback disappointed the children and explaining the point counter colour codes was time-consuming, we did not reuse this exercise.

\subsection{Resources: Programming Environment}
\paragraph{\bf Programming with OzoBlockly.}
Programming the robots with OzoBlockly worked well as the icons are self-explanatory and the students intuitively understood how to drag and drop blocks. Once the complexity and length of programs increased, it turned out that deleting blocks was less self-explanatory and often needed explicit instruction. In Scratch~\cite{maloney2010}, hat blocks can be used to specify at which point in time scripts are executed. OzoBlockly does not have such event handler blocks and in combination with scripts which were not deleted, this led to unexpected robot behaviour, as the order and number of scripts executed is unclear when more than one script is present in the editor. Therefore selection of an appropriate programming language is an inevitable part of the lesson preparation (e.g. the robot Thymio can be programmed with VPL, Scratch, Blockly and Aseba Studio).

\paragraph{\bf Language Barriers.} The OzoBlockly app supports several languages, but the set of languages available is very limited. One of the reasons why we used the pre-reader level for the introduction was indeed that \german is not a supported language. Although we contacted the Ozobot company, their response was that additional languages will be added only once there has been sufficient demand. At the time of this writing, apparently this has not been the case for \german, but we hope that this will change in due time.

\subsection{Students}
\paragraph{\bf Understanding the Repeatable Nature of Programs.}
Some of the children did not intuitively understand the repeatable nature of programs. In Activity 7, the task was to get the robot from a fixed starting point to the house. Some of the children tried to use the editor as remote control by running several different programs until eventually the robot reached the goal from wherever it stopped after execution of the previous program. When moving the robot back to the fixed starting point and running the current program again, the robot did not reach the house. After explicitly hinting at the goal to write one coherent, repeatable program with the robot starting from the fixed starting point, the children usually got it right quickly. This misunderstanding is not specific to programming Ozobots and also occurred in some of our other courses using BeeBots.

\paragraph{\bf Complexity of Programs.}
Using pen-and-paper made it easy for the children to understand how to use colour codes to program the robots. As in Activity~4 they already had to think about the temporal interplay of commands, the transition from drawing to programming with commands seemed easy for them. However, due to time constraints the complexity of the programs the children created with the OzoBlockly editor was limited. 
Most of the time, we did not introduce loops or conditionals. In the cases where we did, the children had created very long scripts with repeating blocks and were eager to learn about a better solution for their code.

\paragraph{\bf Preconceptions against Programming.}
Overall, most of the children were excited to work with robots, and for both boys and girls, being good at programming was something worth striving for. Girls who did a good job at programming were happy to receive positive feedback and be called good programmers. We did not recognise any gender-specific preconceptions against programming or computer science, which resembles the current state of research~\cite{geldreich2019perceptions}. Also, boys and girls were particularly fascinated by one of the robots which we decorated with a unicorn skin. 

\subsection{Teachers}
\paragraph{\bf Preparing Programs for Students.} The erroneous programs used for Activity 7 had to be created manually on every tablet computer, which was a time-consuming process. We strongly recommend to try the newly launched Ozobot classroom platform\footnote{https://ozobot.com/educate/classroom, last accessed June 1, 2021} which claims to solve this problem, but was not yet available during our outreach activities. Alternatively, one could let the students copy the erroneous code from the worksheet themselves, which would also let them
 practice dragging-and-dropping blocks.

\paragraph{\bf Checking Which Codes Are Needed.}
The colour code reference sheet we used in our workshops did not include the code for ``pause'' at the time, even though this is needed for Activity 5. Also, the codes for the counters needed in the ``Modeling animal habits'' lesson were not included. We chose to draw the missing codes on the blackboard. 
Although the current version of the colour codes reference\footnote{https://play.ozobot.com/print/guides/ozobot-ozocodes-reference.pdf, p. 1, last accessed June 1, 2021} contains all of these codes , we recommend checking which codes are supposed to be used for the activities planned, so that one can provide the students with the information they need.

\paragraph{\bf Using an Appropriate Approach.} When working with primary school students, both the pace and the design of the activities have to be adjusted \cite{mayerova2019creating}. Our approach of proceeding step-by-step and involving the children actively has proven to be successful: The often described problems of drawing accurate lines and colour codes \cite{geier2017einsatz,fojtik2017ozobot} hardly occurred. Every pair of children programmed a functioning dance and the learnt concept of sequence could be transferred to BeeBots in some courses in which we had the time to introduce these afterwards. 
Last but not least, all children stated having enjoyed programming with the Ozobot which is one of the central objectives when programming in primary schools.


\subsection{Ozobots in CoderDojos}

Independently of the project underlying this experience report, we regularly organise a CoderDojo\footnote{https://coderdojo.com, last accessed June 1, 2021}, in which children work on their individual projects, with mentors  available to help them. Some of the children wanted to work with the Ozobots. The workshop format we had developed works well with classes where everyone works on the same project, but in our experience it did not work well in the CoderDojo setting. Introducing Ozobots required one of the instructors to be present all the time, as we had no tutorials in \german and the more complex blocks require a level of English skills which not all of the children had. Many ended up having fun with the remote control 
instead of programming the robots.

\section{Conclusions}

Introducing children to programming concepts at primary school level aims at fostering computational thinking skills and overcoming preconceptions and gender imbalance. In order to support this endeavour, we organised programming workshops  at local schools. Due to their size and functionality the Ozobot robots turned out to be well suited for this task. Despite some challenges such as battery life, on the whole our experience was overwhelmingly positive.

Our workshops so far were limited to two hours, and thus only allowed us to teach basic programming concepts; in most cases we covered no more than sequences. 
A particular feature of Ozobot robots is their dual mode of operation in a pen-and-paper mode as well as block-based programming mode. In our experience the pen-and-paper mode is a great way to introduce children to the abilities of the robots, and to build up enthusiasm for robots in general. 
While the relevance of the initial activities in pen-and-paper mode with respect to computer science learning taxonomies~\cite{fuller2007developing} can clearly be argued, it remains unclear whether the concrete skills learned during this mode of operation support the subsequent learning of more advanced concepts related to programming.

Therefore, as a next step we would like to explore the use of Ozobot robots when continuing with introducing further concepts and gradually progressing learners to more advanced programming. With each of the five levels in the OzoBlockly app the complexity increases, and the top level seems equally well suited for older learners. Due to the possibility of viewing a JavaScript representation of OzoBlockly programs, it is even conceivable to use Ozobots throughout the learning process, up to text-based programming.

\vspace{-0.3em}
 \begin{acks}
\vspace{-0.3em}
This work was supported by INTERREG Österreich-Bayern project NB-23 ``Grenzüberschreitende Förderung der Informatikbildung''.
\end{acks}

\bibliographystyle{ACM-Reference-Format}
\bibliography{references} 

\end{document}